\documentclass[a4paper,12pt]{article}
\usepackage[top=1.25in,left=1in,right=1in,bottom=1.25in]{geometry}           
\usepackage{latexsym}
\usepackage{epsfig, ecltree, epic, eepic}
\usepackage{enumerate,amsmath,amssymb,dsfont,pifont,amsthm}
\usepackage[dvips]{color}
\usepackage{graphicx}
\usepackage[arrow, matrix, curve]{xy}

\theoremstyle{plain}
\newtheorem{theorem}{Theorem}[section]

\theoremstyle{definition}
\newtheorem{definition}[theorem]{Definition}

\theoremstyle{remark}
\newtheorem{remark}[theorem]{Remark}
\newtheorem{remarks}[theorem]{Remarks}

\newcommand{\bbr}{\mathbb{R}}

\newcommand{\bbp}{\mathbb{P}}

\newcommand{\bbn}{\mathbb{N}}

\begin{document}

\allowdisplaybreaks

\title{\bfseries An Ordinal Pattern Approach to Detect and to Model Leverage Effects and Dependence Structures Between Financial Time Series}

\author{%
    \textsc{Alexander Schnurr}%
    \thanks{Lehrstuhl IV, Fakult\"at f\"ur Mathematik, Technische Universit\"at Dortmund,
              D-44227 Dortmund, Germany,
              \texttt{alexander.schnurr@math.tu-dortmund.de}}
    }

\date{\today}

\maketitle
\begin{abstract}
We introduce two types of ordinal pattern dependence between time series. 
Positive (resp. negative) ordinal pattern dependence can be seen as a non-paramatric and in particular non-linear counterpart to positive (resp. negative) correlation.
We show in an explorative study that both types of this dependence show up in real world financial data. 
\end{abstract}



\section{Introduction and General Setting}

Back in 1976 Black analyzed the effect that the asset returns and the corresponding volatility are negatively correlated. In the financial literature this phenomenon is referred to as \emph{leverage effect}. In later studies the existence of this effect was affirmed by several authors (see e.g. Yu (2005) and the references given therein). The classical way to capture the leverage effect in models for stock markets is to assume a negative correlation between the two datasets which is constant in time (Heston (1993), Barndorff-Nielsen and Shepard (2002)). However, there is strong evidence that this effect is not constant, but itself evolves in time. It seems that there are periods where the effect is weaker and sometimes it even seems to be `turned around', i.e., there is a positive correlation between the two datasets (Carr and Wu (2007)). Taking these empirical findings into account, more sophisticated models where suggested. The correlation structure was modeled by a deterministic function or was made state space dependent (Romano and Touzi (1997), Bandi and Reno (2009)). In Veraart and Veraart (2010) the correlation structure is itself a stochastic process on $[-1,1]$. Compare in this context also Emmerich (2007).

The current situation is as follows: a vast literature on modeling leverage effects is facing huge datasets of hundreds of assets with the corresponding volatilities. The same problem appears in various parts of mathematical finance: given two datasets $x$ and $y$ and two models (these are $\bbr$-valued stochastic processes) $X$ and $Y$:
\begin{itemize}
  \item Do the models fit the respective datasets? \emph{The} classical question of stochastic modeling.
  \item Is there a dependence structure between $x$ and $y$ and is this modeled correctly by $X$ and $Y$?
  \item What is the nature of the dependence structure? Do we have linear dependence?
\end{itemize}
In the present paper we are concerned mainly with the second question. Instead of proposing even more complicated models we introduce a rather simple approach to analyze whether there is a dependence structure between two datasets. In order to capture the zik-zak of the datasets we use so called ordinal patterns. This method was developed by Bandt (2005) and Keller et al. (2007) in order to handle time series with several thousands of data points which appear in medicine, meteorology and finance (cf. Keller and Sinn (2005)). We compare the two datasets from this point-of-view, that is, we estimate the probabilities of coincident respective reflected patterns (see below). On some occasions, as an example we will consider the S\&P 500 and the corresponding volatility index VIX, a dependence structure of this kind seems to be more likely to be found in real data than the dependence modeled by the classical approach via correlation (cf. Whaley (2008)). 

The basic idea is simple to state: let us have a look at four consecutive data points of the first dataset: $x_n,x_{n+1},x_{n+2},x_{n+3}$.  The second point $x_{n+1}$ can be higher or lower\footnote{We state below how to deal with the case $x_n=x_{n+1}$.} than $x_n$. Let us assume we have $x_{n+1}>x_n$. Then the third point $x_{n+2}$ can be higher than both anterior values, or it can be in between the two or lower then both:

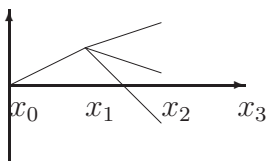
\begin{center}
\setlength{\unitlength}{1cm}
\begin{picture}(1,2)
\put(0,0){\vector(0,1){2}}
\put(0,1){\vector(1,0){3.1}}
\linethickness{0.5mm}
\put(0,1){\line(2,1){1}}
\put(1,1.5){\line(3,1){1}}
\put(1,1.5){\line(3,-1){1}}
\put(1,1.5){\line(3,-3){1}}
\put(0,0.6){$x_0$}
\put(1,0.6){$x_1$}
\put(2,0.6){$x_2$}
\put(3,0.6){$x_3$}
\end{picture}
\end{center}
In each of the three cases we have four possibilities where the fourth value $x_{n+3}$ might be placed: it can be the highest of the four, the second highest and so on.
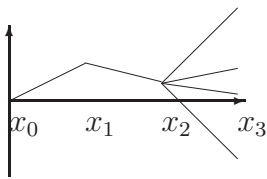
\begin{center}
\setlength{\unitlength}{1cm}
\begin{picture}(1,2)
\put(0,0){\vector(0,1){2}}
\put(0,1){\vector(1,0){3.1}}
\linethickness{0.5mm}
\put(0,1){\line(2,1){1}}
\put(1,1.5){\line(4,-1){1}}
\put(2,1.23){\line(1,1){1}}
\put(2,1.23){\line(5,1){1}}
\put(2,1.23){\line(7,-1){1}}
\put(2,1.23){\line(1,-1){1}}
\put(0,0.6){$x_0$}
\put(1,0.6){$x_1$}
\put(2,0.6){$x_2$}
\put(3,0.6){$x_3$}
\end{picture}
\end{center}
This results in 24 possibilities for the zik-zak of the data points. In order to describe these in a compact way, we write each one of them as a permutation. This is done as follows: write down the values in assorted form (top-to-bottom), say $x_{n+3}>x_{n+1}>x_{n+2}>x_n$. The order of the indices (minus $n$) yields the permutation $(3,1,2,0)$. This permutation is called the \emph{(ordinal) pattern} of the four data points. If it happens that $x_j=x_k$ for a $j<k$ we put $...,j,k,...$ in the permutation. The mapping which associates the permutation with the data points is denoted by $\Pi$. Instead of keeping the full information of the dataset $x$ one now writes down the sequence of patterns. This can be done in two ways: either one moves forward always one point, that is, analyze $(x_0,x_1,x_2,x_3)$, $(x_1,x_2,x_3,x_4)$, ... or the whole block $(x_0,x_1,x_2,x_3)$, $(x_3,x_4,x_5,x_6)$, ... In the first case one might face problems in overestimating dependence structures while in the second one some information might be lost.

The same analysis is carried out for the second dataset $y$. Coming back to the three questions stated above: the probabilities of the ordinal patterns in both time series should be approximately equal to the relative frequency of the patterns in the respective datasets. Furthermore, if there is a dependence structure between the patterns of the two datasets, then this should be the same in the ordinal pattern probabilities of the models $X$ and $Y$. Matching the datasets and certain models or classes of models is part of ongoing research and not in the scope of the present article. Compare in this context Keller and Sinn (2011). Here we just introduce our new concept of dependence and show that there is strong evidence that a dependence structure of this kind exists between real world financial time series. 

An approach via ordinal patterns has the following advantages:
for several classes of processes the ordinal pattern probabilities can be calculated and there are quick algorithms to analyze the relative frequencies of ordinal patterns in given datasets. The dependence analysis is so simple that it can be carried out even by a standard spreadsheet application. 
The whole analysis is stable under monotone transformations and
small perturbations of the data or measure errors do not destroy the ordinal structure.

Let us fix some notation:
the main objects under consideration are two discretely observed stochastic processes $(X_n)_{n\in\bbn_0}$ and $(Y_n)_{n\in\bbn_0}$. For our purpose it does not matter if the processes are a-priori discrete or if a continuous process is observed at times $t_0<t_1<...$. 
In order to keep the notation simple we will always use $\bbn_0$ as index set. The increments are denoted by $(\Delta X_n)_{n\in\bbn}$ where $\Delta X_n:=X_n-X_{n-1}$.
$h\in\bbn$ is the number of consecutive increments under consideration, $\pi$ denotes elements of the symmetric group $S_h$ and
\[
  \bbp_n(\pi):=\bbp(\Pi(X_n,X_{n+1},...,X_{n+h})=\pi).
\]
For an element $\pi\in S_h$, $m(\pi)$ is the `reflected permutation', i.e., read the permutation from right to left.

The paper is organized as follows: in Section 2 we introduce our concept of ordinal pattern dependence between time series. 
In Section 3 we analyze two real world examples. We close the paper by giving an outlook on open questions and further generalizations in Section 4. 

\section{Ordinal Pattern Dependence}

In the following we analyze and model the dependence structure between time series via ordinal pattern probabilities. Let us emphasize that the ordinal pattern probabilities are a property of the increments of the time series. In the case $h=2$ we obtain 
\begin{align} \begin{split} \label{patternprobs}
  \bbp_n((2,1,0))&=\bbp(\Delta X_{n+1}> 0,\Delta X_{n+2}> 0) \\
    \bbp_n((1,2,0))&=\bbp(\Delta X_{n+1}  >  0,\Delta X_{n+2}\leq 0,\Delta X_{n+1}+\Delta X_{n+2} > 0)\\
  \bbp_n((1,0,2))&=\bbp(\Delta X_{n+1}>0,\Delta X_{n+2}\leq 0,\Delta X_{n+1}+\Delta X_{n+2}\leq0) \\  
  \bbp_n((0,1,2))&=\bbp(\Delta X_{n+1}  \leq  0,\Delta X_{n+2}  \leq  0) \\
  \bbp_n((0,2,1))&=\bbp(\Delta X_{n+1} \leq 0,\Delta X_{n+2} > 0,\Delta X_{n+1}+\Delta X_{n+2}\leq 0)\\
    \bbp_n((2,0,1))&=\bbp(\Delta X_{n+1}\leq 0,\Delta X_{n+2}>0,\Delta X_{n+1}+\Delta X_{n+2}> 0) .
\end{split} \end{align}
For several classes of models these probabilities are known or they can (at least) be approximated. Since we want to work model free, we estimate these probabilities using relative frequencies. In our analysis we restrict ourselves most of the time to the cases $h=2$ and $h=3$, that is, we analyze three respectively four consecutive data points. The case of general $h\in\bbn$ can be treated in the same manner. However, in doing this $(h+1)!$ patterns have to be considered. Studying the cases of small values of $h$ is in line with the existing literature (cf. e.g. Keller and Sinn (2011))

\textbf{Assumption:}
We always assume that the time series under consideration are \emph{ordinal pattern stationary}, that is, for every $\pi \in S_h$ the probability $\bbp_n(\pi)$ does not depend on $n$.

\begin{remark}
Obviously stationarity of a time series implies stationary increments, which in turn implies ordinal pattern stationarity. In fact our assumption is just a working hypotheses. Later the ordinal pattern analysis can be used to actually detect structural breaks.	
\end{remark}

In order to describe how strong the dependence between the two time series is, we use the following concept.

\begin{definition}
The time series $X$ and $Y$ exhibit a positive ordinal pattern dependence (ord$\oplus$) of order $h\in\bbn$ and level $\alpha>0$ if
\begin{align*}
 \bbp&\Big(\Pi(X_n,X_{n+1},...,X_{n+h})=\Pi(Y_n,Y_{n+1},...,Y_{n+h})\Big) \\
 &> \alpha + \sum_{\pi\in S_h} \bbp\Big(\Pi(X_n,X_{n+1},...,X_{n+h})=\pi\Big) \cdot \bbp\Big(\Pi(Y_n,Y_{n+1},...,Y_{n+h})=\pi\Big)
\end{align*}
and negative ordinal pattern dependence (ord$\ominus$) of order $h\in\bbn$ and level $\beta>0$ if
\begin{align*}
 \bbp&\Big(\Pi(X_n,X_{n+1},...,X_{n+h})=m\big(\Pi(Y_n,Y_{n+1},...,Y_{n+h})\big)\Big) \\
 &> \beta + \sum_{\pi\in S_h} \bbp\Big(\Pi(X_n,X_{n+1},...,X_{n+h})=\pi\Big) \cdot \bbp\Big(m(\Pi(Y_n,Y_{n+1},...,Y_{n+h}))=\pi\Big).
\end{align*}
where $m(\pi)$ denotes the reflected permutation (see above).
\end{definition}

\begin{remarks}
(a) An ord$\oplus$ means that for a large number of indices $j$ an isotone transformation $\bbr\to\bbr$, mapping $x_k$ to $y_k$ (for $k\in\{j,j+1,...,j+h\}$) exists, where $(x_k)_{k\in \{0,...,N\}}$ is a realization of $X$. If there exists an antitone transformation with the same property, this corresponds to an ord$\ominus$. 

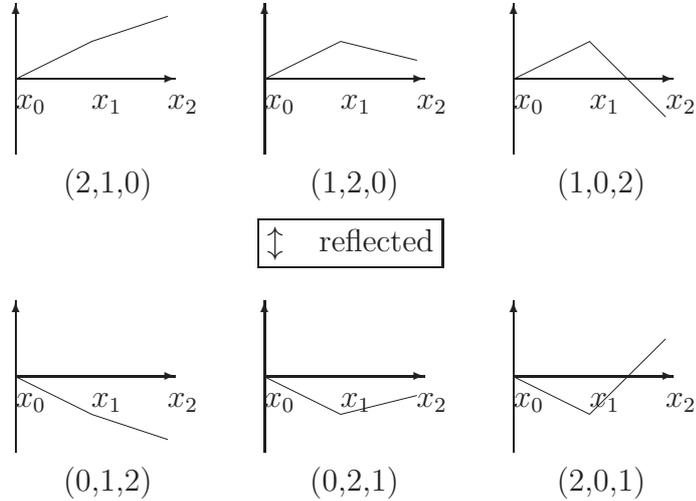
\centerline{
\setlength{\unitlength}{1cm}
\begin{picture}(1,2)
\put(0,0){\vector(0,1){2}}
\put(0,1){\vector(1,0){2.1}}
\linethickness{0.5mm}
\put(0,1){\line(2,1){1}}
\put(1,1.5){\line(3,1){1}}
\put(0,0.6){$x_0$}
\put(1,0.6){$x_1$}
\put(2,0.6){$x_2$}
\end{picture}
\hspace{2cm}
\begin{picture}(1,2)
\put(0,0){\vector(0,1){2}}
\put(0,1){\vector(1,0){2.1}}
\linethickness{0.5mm}
\put(0,1){\line(2,1){1}}
\put(1,1.5){\line(4,-1){1}}
\put(0,0.6){$x_0$}
\put(1,0.6){$x_1$}
\put(2,0.6){$x_2$}
\end{picture}
\hspace{2cm}
\begin{picture}(1,2)
\put(0,0){\vector(0,1){2}}
\put(0,1){\vector(1,0){2.1}}
\linethickness{0.5mm}
\put(0,1){\line(2,1){1}}
\put(1,1.5){\line(1,-1){1}}
\put(0,0.6){$x_0$}
\put(1,0.6){$x_1$}
\put(2,0.6){$x_2$}
\end{picture}
}
\centerline{\hspace{15mm}(2,1,0) \hspace{18mm} (1,2,0) \hspace{18mm} (1,0,2)}

\vspace{2mm}

\centerline{\hspace*{14mm}\framebox{$\updownarrow$ \hspace{2mm} reflected}}

\vspace{4mm}

\centerline{
\setlength{\unitlength}{1cm}
\begin{picture}(1,2)
\put(0,0){\vector(0,1){2}}
\put(0,1){\vector(1,0){2.1}}
\linethickness{0.5mm}
\put(0,1){\line(2,-1){1}}
\put(1,0.5){\line(3,-1){1}}
\put(0,0.6){$x_0$}
\put(1,0.6){$x_1$}
\put(2,0.6){$x_2$}
\end{picture}
\hspace{2cm}
\begin{picture}(1,2)
\put(0,0){\vector(0,1){2}}
\put(0,1){\vector(1,0){2.1}}
\linethickness{0.5mm}
\put(0,1){\line(2,-1){1}}
\put(1,0.5){\line(4,1){1}}
\put(0,0.6){$x_0$}
\put(1,0.6){$x_1$}
\put(2,0.6){$x_2$}
\end{picture}
\hspace{2cm}
\begin{picture}(1,2)
\put(0,0){\vector(0,1){2}}
\put(0,1){\vector(1,0){2.1}}
\linethickness{0.5mm}
\put(0,1){\line(2,-1){1}}
\put(1,0.5){\line(1,1){1}}
\put(0,0.6){$x_0$}
\put(1,0.6){$x_1$}
\put(2,0.6){$x_2$}
\end{picture}
}

\centerline{\hspace{15mm}(0,1,2) \hspace{18mm} (0,2,1) \hspace{18mm} (2,0,1)}
(b) By that definition we can obviously have positive and negative dependence the same time. At least if $\alpha$ and $\beta$ are not too big.

(c) From a statistical point-of-view this means that if we consider real data, we are testing against the hypothesis `independence' and the critical region is $]\alpha,1]$ respective $]\beta,1]$. In our analysis we estimate the ordinal pattern probabilities of both time series, using the given data. This allows us to work model free.

(d) Besides the classical concept of negative correlation, there are other concepts of negative dependence in the literature: Zarei and Jabbari (2011) [Definition 1.1] and Sung (2012) use a different concept of negative dependence \emph{within one} time series, which was first introduced by Lehmann (1966). 

\end{remarks}


\section{Two Empirical Examples}

As a first real-world example we consider the dependence structure between the well known S\&P 500 index (SPX) and the corresponding Chicago Board Options Exchange Volatility Index (VIX), which was introduced by R. E. Whaley in 1993 and which was revised in 2003. The VIX is used here as an empirical measure for the volatility of the underlying index (compare in this context Nagel and Sch\"obel (1999) who use the VDAX for the same purpose). The interplay between these two time series is a vital part of ongoing research (cf. Madan and Yor (2011)) and it is interesting from our point-of-view for the following reason: in the standard literature one finds that they are negatively correlated. However, the inventor of the index states in his paper from 2008 the fact that if the SPX increases, the VIX decreases and the other way around but `the relation between rates of change in the VIX and the SPX is asymmetric' (page 7), in particular there is no linear dependence, which would be modeled by correlation. 

The second example we analyze deals with the dependence between the Dow Jones and the Dollar-Euro exchange rate. There is strong evidence that there is a `positive dependence' between these two empirical time series. We show that there seems to be a positive ordinal pattern dependence in the time period we have analyzed. 

\subsection{Data} In both examples we have used open source data. We have extracted the historical data for SPX, VIX and Dow Jones  from finance.yahoo.com and the source for the Dollar-Euro exchange rate has been www.global-view.com/forex-trading-tools. In case of the three indices we have used the `close prices', if not mentioned otherwise. In our second comparison we had to erase the data of the Euro-Dollar exchange on dates which are public holidays in the USA in order to make the datasets comparable. 

Using the first approach described in the introduction, that is, we have moved forward point-by-point, we have chosen 503 consecutive data points. The first one is not used in the case $h=2$. Therefore, we have 500 patterns in both cases. Using the second approach (blockwise), we have chosen 1002 data points in the case $h=2$ respectively 1503 data points in the case $h=3$. This results again in 500 patterns, making it easy to compare the results. Since the data of the VIX only goes back to 1990, in analyzing monthly data we had to use less data points: we restricted ourselves to 250 patterns.

\subsection{Method of Analysis}

We have analyzed the datasets described above using a spreadsheet application first and then a program written in R. Using the spreadsheet application we assigned a number from 0 to 5 ($h=2$) respectively 0 to 23 ($h=3$) to every pattern in such a way that adding the numbers of reflected patterns yields 5 respectively 23. 

In order to estimate the ord$\oplus$ respectively the ord$\ominus$, we need the following quantities:

\begin{definition}
We write
\begin{align*}
 &p_{eq} & &\bbp\Big(\Pi(X_n,X_{n+1},...,X_{n+h})=\Pi(Y_n,Y_{n+1},...,Y_{n+h})\Big) \\
 &p_{neq}& \text{ for the estimate of } \hspace{5mm}&\bbp \Big(\Pi(X_n,X_{n+1},...,X_{n+h})=m\big(\Pi(Y_n,Y_{n+1},...,Y_{n+h})\big)\Big) \\
 &p_\pi^X& & \bbp(\Pi(X_n,X_{n+1},...,X_{n+h})=\pi) \text{ for } \pi\in S_h.
\end{align*}
via relative frequencies, and 
\begin{align*}
\widetilde{\alpha}&:= p_{eq}- \sum_{\pi\in S_h} (p_\pi^X \cdot p_\pi^Y) \\
\widetilde{\beta}&:= p_{neq}- \sum_{\pi\in S_h} (p_\pi^X \cdot p_{m(\pi)}^Y)
\end{align*}
for the model free estimates of the ord$\oplus$level $\alpha$ and the ord$\ominus$level $\beta$.
\end{definition}

\subsection{Results}

\emph{i) a detailed first example}

In order to have a first example we have chosen the time period 27/01/2010 to 24/01/2012 randomly. We have analyzed daily data, moving forward one-by-one. Our main interest has been the  dependence structure between the SPX and the VIX respectively between the Dow Jones and the Dollar-Euro exchange rate.

For the first example (SPX/VIX) we present the detailed results in the following table. Similar tables have been produced for every analyzed dataset.
\vspace*{4mm}

\centerline{
\begin{tabular}{|c|c|c|c|c|c|c|}
pattern   & abs.freq & abs.freq & rel.freq & rel.freq  &$p_\pi^X \cdot p_\pi^Y$ &  $ p_\pi^X \cdot p_{m(\pi)}^Y$ \\ 
   $\pi$       & SPX & VIX      & SPX & VIX & & \\ \hline
(0,1,2,3) & 45 & 91 & 0.090 & 0.182 & 0.016380 & 0.009180 \\ \hline 
(0,1,3,2) & 24 & 37 & 0.048 & 0.074 & 0.003552 & 0.002400 \\ \hline
(0,3,1,2) & 27 & 19 & 0.054 & 0.038 & 0.002052 & 0.001836 \\ \hline
(3,0,1,2) & 10 & 15 & 0.020 & 0.030 & 0.000600 & 0.000640 \\ \hline
(0,2,1,3) & 11 & 14 & 0.022 & 0.028 & 0.000616 & 0.000836 \\ \hline
(0,2,3,1) & 07 & 12 & 0.014 & 0.024 & 0.000336 & 0.000196 \\ \hline
(0,3,2,1) & 15 & 13 & 0.030 & 0.026 & 0.000780 & 0.000960 \\ \hline
(3,0,2,1) & 17 & 18 & 0.034 & 0.036 & 0.001224 & 0.001088 \\ \hline
(2,0,1,3) & 11 & 09 & 0.022 & 0.018 & 0.000396 & 0.000220 \\ \hline
(2,0,3,1) & 10 & 07 & 0.020 & 0.014 & 0.000280 & 0.000080 \\ \hline
(2,3,0,1) & 08 & 14 & 0.016 & 0.028 & 0.000448 & 0.000320 \\ \hline
(3,2,0,1) & 38 & 28 & 0.078 & 0.056 & 0.004256 & 0.005928 \\ \hline
(1,0,2,3) & 32 & 39 & 0.064 & 0.078 & 0.004992 & 0.003584 \\ \hline
(1,0,3,2) & 12 & 10 & 0.024 & 0.020 & 0.000480 & 0.000672 \\ \hline
(1,3,0,2) & 08 & 02 & 0.016 & 0.004 & 0.000064 & 0.000224 \\ \hline
(3,1,0,2) & 12 & 05 & 0.024 & 0.010 & 0.000240 & 0.000432 \\ \hline
(1,2,0,3) & 18 & 16 & 0.036 & 0.032 & 0.001152 & 0.001296 \\ \hline
(1,2,3,0) & 11 & 16 & 0.022 & 0.032 & 0.000704 & 0.000572 \\ \hline
(1,3,2,0) & 06 & 07 & 0.012 & 0.014 & 0.000168 & 0.000288 \\ \hline
(3,1,2,0) & 18 & 19 & 0.036 & 0.038 & 0.001368 & 0.001008 \\ \hline
(2,1,0,3) & 19 & 16 & 0.038 & 0.032 & 0.001216 & 0.001140 \\ \hline
(2,1,3,0) & 23 & 17 & 0.046 & 0.034 & 0.001564 & 0.001748 \\ \hline
(2,3,1,0) & 28 & 25 & 0.056 & 0.050 & 0.002800 & 0.004144 \\ \hline
(3,2,1,0) & 90 & 51 & 0.180 & 0.102 & 0.018360 & 0.032760 \\ \hline
\end{tabular}
}

In the time period described above we have found 241 \emph{reflected patterns} (out of 500) in the case $h=2$ and 144 in the case $h=3$. The estimate for the ord$\ominus$ is $\widetilde{\beta}=0.2164$, that is, the estimated probability of reflected patterns is more than 0.2 higher than one would suspect under the hypothesis of independence! In order to emphasize the differences in the dependence structure we compare these findings with the numbers for coincident patterns: in the case $h=2$ we found 21 and in the case $h=3$ only 5. 

In the same period of time we found the following for the Dow Jones compared with the Dollar-Euro exchange: in the case $h=2$ there were 192 out of 500 pattern which did \emph{coincide} and in the case $h=3$ still 93 out of 500. Again we compare this with the reflected patterns; here we have obtained 41 respectively 10. The estimate for the ord$\oplus$ is $\widetilde{\alpha}=0.186-0.0636=0.1224$.


\emph{ii) daily data in time}

We consider again the first example (SPX/VIX) and analyze how the frequencies of (some) ordinal patterns ($h=3$) and the dependence structure changes over time. To this end we plot the following table:

\vspace*{4mm}
\centerline{
\begin{tabular}{|c|c|c|c|c|c|c|c|c|c|}
from  & to & h=3 & h=2& $\pi_0$ & $\pi_0$ & $\pi_6$ & $\pi_6$ & $\pi_{12}$ & $\pi_{12}$\\
  & & refl & refl& SPX & VIX & SPX & VIX & SPX & VIX \\ \hline
12/05/2010 & 08/05/2012 & 176 & 279 & 83 & 46 & 14 & 14 & 38 & 35 \\ \hline 
13/05/2008 & 11/05/2010 & 196 & 306 & 71 & 31 & 12 & 16 & 34 & 32 \\ \hline
12/05/2006 & 12/05/2008 & 197 & 297 & 64 & 40 & 18 & 18 & 32 & 19 \\ \hline
13/05/2004 & 11/05/2006 & 163 & 268 & 75 & 31 & 11 & 15 & 29 & 21 \\ \hline
15/05/2002 & 12/05/2004 & 139 & 253 & 54 & 53 & 12 & 19 & 30 & 31 \\ \hline
10/05/2000 & 14/05/2002 & 179 & 281 & 43 & 51 & 18 & 17 & 34 & 33 \\ \hline
13/05/1998 & 09/05/2000 & 212 & 318 & 66 & 48 & 10 & 12 & 42 & 44 \\ \hline
13/05/1996 & 12/05/1998 & 150 & 258 & 78 & 66 & 14 & 14 & 47 & 29 \\ \hline
17/05/1994 & 10/05/1996 &  97 & 184 &106 & 64 & 14 & 18 & 37 & 26 \\ \hline
20/05/1992 & 16/05/1994 & 107 & 202 & 73 & 50 & 15 & 11 & 41 & 30 \\ \hline
23/05/1990 & 19/05/1992 & 110 & 210 & 54 & 40 & 13 & 17 & 38 & 33 \\ \hline 
\end{tabular}}

Here $\pi_0=(0,1,2,3), \ \pi_6=(0,3,2,1)$ and $ \pi_{12}=(1,0,2,3)$. 

\emph{iii) weekly and monthly data}

We compare the daily data (SPX/VIX) with weekly and monthly data:

For weekly data in the time period 23/09/2002 to 07/05/2012 analyzed one-by-one we get in the case $h=2$ still 254 reflected patterns and in the case $h=3$ we get 148.

Being invented in 1993 the VIX has been calculated back until 1990. This means it exists by now for less than 500 month. For monthly data we restrict ourselves to 253 data points, resulting in 250 patterns. In the time period 01/04/1991 to 01/05/2012 analyzed one-by-one we get 113 reflected patterns in the case $h=2$ and 50 in the case $h=3$.

\emph{iv) open vs close data}

Above we have always considered the `close data' of the prices for $h=2$ respectively $h=3$. If we consider the `open data' instead we obtain in the case of daily data (12/05/2010 to 08/05/2012) 203 out of 500 respectively 110,
and in the case of weekly data (23/09/2002 to 07/05/2012) 228 out of 500 respectively 124.
Within the same periods of time we obtain for the `high data' (`low data') evaluating daily data: 218 (253) respectively 116 (152) reflected patterns and for weekly data 199 (174) respectively 100 (78).

\emph{v) datasets without ordinal pattern dependence}

We compare our above findings with two other examples:\\
Take the open and the close data of the weekly VIX-data in the time period 23/09/2002 to 07/05/2012. We obtain for $h=2$: 102 coincident patterns ($\widetilde{\alpha}=0.03$) and for $h=3$: 39 coincident patterns ($\widetilde{\alpha}=0.02$). \\
Simulating two independent datasets with standard-normal increments we obtain 101 reflected patterns and 82 coincident patterns in the case $h=2$ and 27 reflected patterns and 25 coincident ones in the case $h=3$. 


\emph{vi) delay}

In our standard example (SPX/VIX, analyzed one-by-one, close data, daily, from 12/05/2010 to 08/05/2012) we have found 279 reflected patterns in the case $h=2$ and 176 in the case $h=3$. Now we shift one of the time-series by one, i.e., we compare the above dataset for the VIX with the dataset 11/05/2012 to 07/05/2012 of the SPX. This results in 106 respectively 40 reflected patterns. Shifting the same set one day into the other direction yields 116 respectively 47 reflected patterns. 

\emph{vii) blockwise analysis} 

Using the second approach mentioned above, that is, moving forward blockwise, we have analyzed the time periods 16/05/2008 to 08/05/2012 for $h=2$ and 22/05/2006 to 08/05/2012 for $h=3$. The datasets under consideration are again SPX and VIX. We obtain 293 reflected patterns for $h=2$ and 189 for $h=3$.

\emph{viii) bigger values of $h$}

We consider the time period 02/01/1990 to 20/12/2012 of the data sets SPX and VIX, that is, we consider 5791 data points of both time series:

\vspace*{4mm}
\centerline{
\begin{tabular}{|c|c|c|c|c|c|c|}
 \hline 
 $h=$ & 1 & 2 & 3 & 4 & 5 & 6 \\ \hline 
coincident patterns & 1303 & 216 & 37 & 5 & 0 & 0 \\ \hline
reflected patterns & 4487& 2992 & 1788 & 987 & 523 & 256  \\ \hline
\end{tabular}}

\subsection{Conclusions}

\emph{a) Existence of ordinal pattern dependence structures}

The most important result for our future research is, that there is strong evidence for the existence of a positive ordinal pattern dependence between the Dow Jones and the Dollar-Euro exchange rate and a negative ordinal pattern dependence between the SPX and the VIX. This means that both concepts, we have introduced, can be found in real world time series. 

\emph{b) Change of the ordinal pattern probabilities over time}

There is no evidence, that the ordinal pattern probabilities change significantly over time. At least as a working hypothesis we can consider the four time series described above as ordinal pattern stationary. Possible exceptions are the probabilities of the patterns (0,1,2,3) and (3,2,1,0). They seem to depend on time.

\emph{c) Change in the dependence structure over time} 

There is strong evidence that the dependence structure changes over time:
in the time period 13/05/1998 to 09/05/2000 we found e.g. 212 reflected patterns in the case $h=3$ and in the time period 17/05/1994 to 10/05/1996 only 97. This is in line with the mathematical finance literature on \emph{linear} dependence: Wilmott (2000) states that `correlations are even more unstable than volatilities'. 

\emph{d) Different time lag (daily, weekly, monthly)}

There is strong evidence that in all of the three analyzed time scales there is a negative ordinal pattern dependence. The estimates for $\beta$ in the weekly and monthly data are lower than the average of the estimates for $\beta$ using daily data in the same period of time. However, they are still highly significant. 

\emph{e) Move forward block-wise}

There is no significant difference between the point-by-point and the blockwise analysis. In fact, we have the following: Consider the data from May 2006 to May 2012 as above. In the three sets consisting of 500 patterns we have found 197,196 resp. 176 reflected patterns. A blockwise analysis yields 189 reflected patterns and the arithmetic mean of 197,196 and 176 is $189.\overline{6}$. 

\emph{f) Delay}

On some occasions one might find delayed effects, that is, an ord$\oplus$/ord$\ominus$ between shifted time-series. This is not the case in the examples we have considered here. A shift of only one day destroys the whole dependence structure.

\emph{g) Different types of dependence between the same time series}

Whenever there was strong evidence for an ord$\ominus$, there was absolutely no evidence for an ord$\oplus$, and the other way around. In fact, if $\widetilde{\alpha}>0$, then $\widetilde{\beta}<0$ and if $\widetilde{\beta}>0$ then $\widetilde{\alpha}<0$. In the time series we have considered, there was no evidence that the leverage effect (in our sense) can be turned around.

\emph{h) Different time windows}

There is strong evidence for an ord$\ominus$ between SPX/VIX for every $h\in\{1,...,6\}$. Even in the case $h=6$, where 5040 patterns have to be considered, 256 out of 5785 are reflected. This should be compared to two independently simulated time series with normal increments. For those we have found in the case $h=6$ only 4 reflected patterns out of 5785 observed patterns. 

\emph{i) Advantages of the method}

Compared to other ways of modeling and analyzing dependence structures, the ordinal pattern dependence has some advantages: as we have mentioned above, the analysis is quick and robust. The interpretation of results is simple and natural. We are able to detect non-linear and asymmetric dependence structures. Unlike correlation, our method is robust against a small number of big measure errors.

\subsection{Final Remarks}

The concepts of (positive/negative) ordinal pattern dependence do appear in real world time series. The effect seems to vary over time. Multivariate stochastic processes which are designed to model such datasets should incorporate ordinal pattern dependence and the possibility that the dependence structure varies over time. On large scales the effect seems to be scale invariant while a delay is not present. 


\section{Outlook and Further Research}

As we have mentioned before, this paper is an explorative study. In the future a careful statistical analysis of several datasets will be carried out in order to affirm or to disprove our above conclusions. Further examples should be considered. In particular it will be of interest whether there are really time series between which the dependence structure can be turned around.

In our analysis, there was no effect in working blockwise instead of point-by-point. This might change if we consider larger values of $h$. For $h$ large it can happen that, because we have a match for $n,n+1,...,n+h$, the probability for a match in $n+1,...,n+h+1$ is much higher.

In the SPX/VIX-example we have seen that the ordinal pattern dependence is scale invariant, at least on large time scales. High frequency data is a different story: in the case of tick-by-tick financial data we face the problem that very often the value does not change at all from one point to the next. By our above convention we will get in the case $h=2$ relatively often the patterns (2,0,1) and (1,2,0) and very often (0,1,2). On the other hand, if there actually exists a delayed effect, it should be in this high frequency regime. If this should be the case, a change in the behavior of one time series can be used to predict the change of the other time series.

Another aspect which we are working on is the relation between ord$\ominus$/ord$\oplus$ and other possibilities to describe and analyze dependence structures. As we have already mentioned above, it is also part of ongoing research to use the method described here in order to match mathematical models with given real world time series. The models we have in mind are defined in continuous time, in order not to have to chose a time scale a priori. It is not in the scope of the present article to go into the details of these model classes. However, let us give a short indication in the direction of modeling: we have succeeded in matching two correlated AR(1)-time series to the given ordinal pattern dependence structure between the SNP 500 and the VIX. We do not care about the scaling and let, therefore, $(Z_n)_{n\in\bbn}$ and $(W_n)_{n\in\bbn}$ be two N(0,1)-iid white noise sequences with $Cor(Z_n,W_n)=-0.8$ for every $n\in\bbn$. Furthermore, we set $\phi:=0.99$, $X_1:=Z_1$, $Y_1:=W_1$,
\[
X_n:=\phi X_{n-1}+Z_n \text{ and } Y_n:=\phi Y_{n-1}+W_n 
\]
for $n\geq 2$. We obtain the following table
\vspace*{4mm}

\centerline{
\begin{tabular}{|c|c|c|c|c|c|c|}
 \hline 
 $h=$ & 1 & 2 & 3 & 4 & 5 & 6 \\ \hline 
coincident patterns & 1205 & 180 & 28 & 3 & 0 & 0 \\ \hline
reflected patterns & 4586& 3063 & 1832 & 1021 & 522 & 255  \\ \hline
\end{tabular}}
which should be compared to the table in Section 3.3 viii) above. In this model the dependence structure is constant in time. It is interesting, that in the case of AR(1)-time series, the concepts of classical negative correlation and ord$\ominus$ are closely related. This is not always the case as the following example emphasizes: simulate two independent N(0,1)-iid sequences of the length 503: $x_1,...,x_{503}$ and $y_1,...,y_{503}$. Now choose randomly 12 points $n_j$ in \{1,...,503 \}. Set $x_{n_j}:=10$ and $y_{n_j}:=-10$ for every $j=1,...,12$. Having simulated this, we got an empirical correlation between the increments of the time series of -0.6937 while we found for $n=2$ only 79 and for $n=3$ only 27 reflected patterns. This means that a small number of correlated measure errors results in a high negative correlation, while it does not have a significant impact on the ord$\ominus$. 

\textbf{Acknowledgements:} The author wishes to thank two anonymous referees for their work. Their comments have helped to improve the paper. Furthermore he wishes to thank B. Funke (TU Dortmund) for carefully reading the manuscript and A. D\"urre (TU Dortmund) for the implementation in R. The financial support of the DFG (German science Foundation) SFB 823: Statistical modeling of nonlinear dynamic processes (project C5) is gratefully acknowledged.  


\end{document}